\documentclass[doublecol,superscriptaddress,floats,floatfix,showpacs]{epl2}
\usepackage{multirow,amssymb,amsbsy,amsmath}
\usepackage{graphicx}
\usepackage{verbatim}
\usepackage{booktabs}
\usepackage{dcolumn}
\usepackage{bbm}
\makeatletter
\def\bra#1{\mathinner{\langle{#1}|}}
\def\ket#1{\mathinner{|{#1}\rangle}}

\DeclareMathOperator{\tr}{tr}

\usepackage{color}

\title{Non-Markovian quantum dynamics: What is it good for?}

\author{C.-F. Li \inst{1,2}$\footnote{email: cfli@ustc.edu.cn}$, G.-C. Guo \inst{1,2} \and
J. Piilo\inst{3}$\footnote{email: jyrki.piilo@utu.fi}$
}

\institute{
  \inst{1} CAS Key Laboratory of Quantum Information, University of Science and Technology of China, Hefei 230026, China\\
  \inst{2} CAS Center For Excellence in Quantum Information and Quantum Physics, University of Science and Technology of China, Hefei 230026, China\\
  \inst{3} Turku Centre for Quantum Physics, Department of Physics and Astronomy, University of Turku, FI-20014 Turun yliopisto, Finland \\
 
  }

\date{\today}

\abstract{
Recent developments in practical quantum engineering and control techniques have allowed significant developments for experimental studies of open quantum systems and decoherence engineering.
Indeed, it has become possible to test experimentally various theoretical, mathematical, and physical concepts related to non-Markovian quantum dynamics. This
includes experimental characterization and quantification of non-Markovian memory effects and proof-of-principle demonstrations how to use them for certain quantum communication and information tasks. We describe here recent experimental advances  for open system studies, focussing in particular to non-Markovian dynamics including the applications of memory effects, and discuss the possibilities for ultimate control of decoherence and open system dynamics.}


\begin{document}
\maketitle

\section{Introduction}
Closed quantum systems evolve by unitary dynamics conserving, e.g.,  the purity of the evolving quantum state.
Thereby, having ability to prepare any initial pure quantum state for a given system, and having arbitrary time dependent and controllable Hamiltonian driving the system, allows to create any pure state trajectory in the system's Hilbert space - at least in principle. For an open quantum system interacting with its environment, the problematics of quantum control becomes  more challenging.
An open system, whose Hamiltonian we may be able to control to certain extent, interacts with a large number of uncontrollable degrees of freedom.  
The open system dynamics becomes non-unitary and the system-environment interaction tends, e.g., to reduce the purity of the open system state via decoherence, i.e., making it more mixed and classical-like. The question now becomes how can we control, not only how decoherence influences the open system dynamics, but also whether it is possible to control the loss of purity -- and even increase the purity and revive the quantum properties at least temporarily?

There are a number of {\it {a priori}} ways one could think of influencing the open system dynamics, indirectly. Since open system evolution depends on how the open system interacts with its environment, then changing the properties of the corresponding interaction Hamiltonian  may become useful. However, for many physical systems the form of the interaction is fixed as soon as we fix the physical system being considered, and therefore have only very limited possibilities to control open system dynamics. More promising avenues may be provided by reservoir engineering, where one controls and manipulates the environment of an open system, i.e., the properties of the physical entity or degrees of freedom with whom the system of interest is interacting with. 
Indeed, one of the first experiments on reservoir engineering was implemented about 20 years ago by applying controlled electric noise on the electrodes of a single ion trap~\cite{reng}.
 It is also worth mentioning  that decoherence and loss of quantum properties often occur in very fast time scale making it difficult to witness in real-time the  loss of quantum properties -- though this is possible requiring very sophisticated experimental set-ups~\cite{sdel}. Note that there also exists implementations of quantum simulators for Markovian open systems motivated by many-body aspects~\cite{jtbar,pschin}.

This Perspective article describes recent experimental developments of decoherence engineering and illustrates how this has opened new possibilities to study non-Markovian quantum dynamics  -- a vividly discussed topic  during the last ten years~\cite{RevRivas,RevRMP,idv,LiRev,per1}.
So far, used experimental platforms include, e.g.,  photons~\cite{NMNP,nmepl,achiu,bhliu13,fff,gyx,bhliu14,Ber2015,sdc,scia1,syu,zdliu,acu,sau}, NMR-systems~\cite{Ber2016,dku}, trapped ions~\cite{mwit}, and nitrogen-vacancy (NV) centers in diamonds~\cite{haa,DJ}.
There exists at least three-fold motivation for these experimental activities i) Elementary control of decoherence dynamics and Markovian to non-Markovian transition ii) Applications of non-Markovian memory effects for quantum information purposes iii) Controlling open system dynamics beyond Markovian to non-Markovian  transition and ultimate limits of decoherence control. We discuss all of these  aspects describing a number of corresponding example  experiments. The current article can be used as a primer for  experimental implementation of non-Markovian quantum dynamics.
For theoretical developments on defining and quantifying non-Markovianity, see the associated  Perspective~\cite{per1} or wide reviews~\cite{RevRivas,RevRMP,idv,LiRev}.

\section{Experimental realizations and control of non-Markovian dynamics}
Controlling the open system and decoherence dynamics in non-Markovian regime can be achieved, e.g., via reservoir engineering or simulating the influence of structured environments. 
By exploiting these concepts, the examples below demonstrate one of the first experiments for controlled Markovian to non-Markovian transition, and the corresponding system-environment information flow control~\cite{NMNP}, the detection of weak to strong non-Markovian transition~\cite{Ber2015} -- both with photons -- followed by a NV-center experiment including an ambient environment~\cite{haa}. 

\subsection{Markovian to non-Markovian transition}
Single photons with polarization and frequency degrees of freedom provide a highly controllable systems where polarization acts as a qubit and frequency as environment~\cite{NMNP}. 
Consider initial factorized polarization-frequency state of a photon,
 $|\Psi(0)\rangle =
 \left(C_H |H\rangle + C_V |V\rangle \right) \otimes \int d\omega g(\omega) |\omega\rangle$, where $C_H$ ($C_V$) is the probability amplitude for polarization $H$ ($V$)
 and $ g(\omega)$ is the probability amplitude for angular frequency $\omega$.
 The interaction between polarization and frequency is provided by birefringent effects -- experimentally implemented, e.g., with quartz plates and described by a Hamiltonian 
 \begin{equation}
 \label{eq:Hp}
 H = (n_H\ket{H}\bra{H} + n_V\ket{V}\bra{V})  \otimes \int  \omega\ket{\omega}\bra{\omega}d\omega,
 \end{equation}
 where $n_H$ ($n_V$) is the index of refraction for polarization component H (V).   The corresponding time evolution operator for polarization direction $\lambda$ and frequency component $\omega$ is now 
 $U(t)|\lambda\rangle \otimes |\omega\rangle
 = e^{in_{\lambda}\omega t} |\lambda\rangle \otimes |\omega\rangle$, i.e.,  the accumulated phases for the probability amplitudes depends on the index of refraction $n_{\lambda}$ of the corresponding polarization and the value of the frequency $\omega$.  By tracing out the frequency degree of freedom, one obtains the dephasing dynamics for the polarization qubit, where the diagonal elements (probabilities) of the qubit density matrix remain constant. The time evolution of the off-diagonal elements, i.e. coherences, are obtained by multiplying their initial values with the decoherence function $\kappa(t)$ as  $\kappa^*(t) C_H^*  C_V |H\rangle\langle V|$ and 
$ \kappa(t) C_H  C_V^* |V\rangle\langle H|$.
The  decoherence function giving the time evolution is directly given by the Fourier transformation of the initual frequency probability  
distribution $| g(\omega) |^2$ 
\begin{equation}
 \kappa(t) = \int d\omega | g(\omega) |^2 e^{i\omega\Delta n t},
\end{equation}
where $\Delta n = n_V-n_H$. This demonstrates that controlling the probability distribution  $| g(\omega) |^2$ allows to control the character of  dephasing. 
In particular, consider $| g(\omega) |^2$   as a sum of two Gaussians each having width $\sigma$ and the central peaks having separation $\Delta \omega$. Now, the magnitude of the decoherence function is given by
\begin{equation}
|\kappa(t)| =
 \frac{e^{-\frac{1}{2}\sigma^2(\Delta nt)^2}}{1+A_{\theta}}
 \sqrt{1+A_{\theta}^2+2A_{\theta}
 \cos(\Delta\omega \cdot \Delta n t)},
 \end{equation}
 where $A_{\theta}$
 parametrizes the relative heights $A_1$ and $A_2$ of the peaks in the following way:  
 $A_1 = \frac{1}{1+A_{\theta}}$, $A_2 = \frac{A_{\theta}}{1+A_{\theta}}$. Having only one frequency peak, $A_{\theta}=0$, the oscillatory part vanishes and one has exponential damping of the magnitude of coherences corresponding to Markovian dynamics. Having equally weighted peaks, $A_{\theta}=1$ and  $\Delta \omega > \sigma$, one has damped oscillatory dynamics of the magnitude of coherences corresponding to non-Markovian dynamics. Thereby controlling the relative heights of the two peaks allows to control the Markovian to  non-Markovian transition. The quantification of amount of non-Markovianity, associated in this case to non-monotonic behaviour of magnitude of coherences, can now be done, e.g., via trace distance measure for non-Markovianity~\cite{NMprl,NMpra}.
 
\begin{figure}[tb]
\begin{center}
\includegraphics[width=0.35\textwidth]{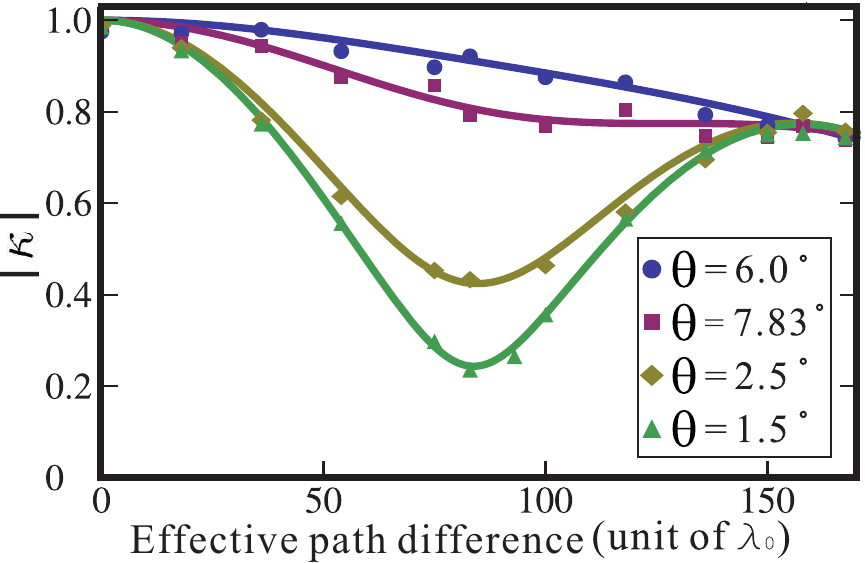}
\caption{\label{fig:1}
The magnitude of the decoherence function $|\kappa(t)|$ as a function of time (measured as effective path difference) for four different values of the tilting angle $\theta$. The symbols correspond to experimental results and solid lines to theoretical fits. Adapted from~\cite{NMNP}.
}
\end{center}
\end{figure}

In the experiment, the photon frequency distribution is engineered -- from single to double peak Gaussian structure -- by filters and inserting a Fabry-Perot cavity on the path of the photon before dephasing begins.
The rotation of the cavity with angle $\theta$ changes the effective thickness of the cavity, that the photon sees, 
and in combination with filters allows precise control of the double peak frequency structure and the relative heights of the two peaks. Figure~\ref{fig:1} shows experimental and theoretical curves of the magnitude of coherences as a function of time  for different tilt angles $\theta$. 
Values $\theta=1.5^{\circ}, 2.5^{\circ}$ correspond to non-Markovian dynamics and $\theta=6.0^{\circ}, 7.83^{\circ}$ to Markovian dynamics.
For more detailed analysis, see~\cite{NMNP}.
 
\subsection{Transition from weak to strong non-Markovianity}

Non-Markovian open systems have rich dynamical features and -- in general -- there exists a large number of quantifiers for non-Markovianity of quantum dynamics~\cite{NMprl,wolf,rivas,channelcap,mani,BD2016,CMM2016,BJA2017}.  This also allows to introduce the concepts of weak and strong non-Markovianity associated to positivity (P) and complete positivity (CP) properties of dynamical maps~\cite{mani}.  
Formally, the time evolution of open system is given by a family of CP dynamical maps $\Phi_{t,0}$ which connects the
initial state of the open system $\rho_S(0)$ to its  evolved state $\rho_S(t)$ at time $t$  as 
$\rho_S(t) = \Phi_{t,0} \rho_S(0)$~\cite{breuer2007}. Considering intermediate points of time  
$t_2\geqslant t_1 \geqslant 0$, one can formally write the map as concatenation 
$\Phi_{t_2,0} =  \Phi_{t_2,t_1} \Phi_{t_1,t_0}$. When the intermediate map $\Phi_{t_2,t_1}$ is both CP and P, the dynamics is classified as Markovian according to this criterion.
However, if the  $\Phi_{t_2,t_1}$ is not P (P but not-CP) then the corresponding dynamics is classified as strongly (weakly) non-Markovian. 

The simplest open system where this transition can be observed is a qubit and
recently photons were used in the corresponding experimental observation~\cite{Ber2015}. 
The implementation is based on a collision-type simulator for open system dynamics where the polarization qubit of the photon goes through two consecutive collisions. Within each of the collisions the evolution consists of probabilistically choosing one of the two Pauli operators $\sigma_x$ or $\sigma_y$ -- or identity operator $\mathbb{I}$.
Let us denote the corresponding probabilities in the first collision with $p_x$, $p_z$, and $p_I.$
The corresponding evolution and the dynamical map is then 
$\rho_S(t_1)= \Phi_{t_1,0} (\rho_S(0)) = p_0 \rho_S(0) + p_x \sigma_x  \rho_S(0)  \sigma_x+ p_z \sigma_z   \rho_S(0) \sigma_z$.
After the second collision, the qubit state can be written in the following way
\begin{equation}
\label{eq:coll}
\rho_S(t_2) = \Phi_{t_2,0} (\rho_S(0)) = \sum_{ij} p_{ij}O_j O_i \rho_S(0)O_i O_j. 
\end{equation}
The summation contains all the possible combinations of operators $\sigma_x$,  $\sigma_z$, and $\mathbb{I}$ with $p_{ij}$ corresponding to  
probability of the joint sequence of the two operators $O_jO_i$.
In the case that the probabilities factorize, $p_{ij}= p_ip_j$, the corresponding collisions are fully independent and the corresponding dynamical map describes Markovian evolution.  In the other extreme, where the second collision is fully dependent on the previous one, the dynamics displays strong non-Markovian memory effects. Consider the following choice for the probabilities:
  $p_{00}=(1-2\epsilon)^2$, $p_{0x}=p_{0z}=p_{z0}=p_{x0}=(1-2\epsilon)\epsilon, p_{xz}=p_{zx}=0$, and $p_{xx}=p_{zz}=2\epsilon^2$ 
 -- and use $\epsilon$ as a control parameter. With these choices, the qubit always experiences non-Markovian evolution and by the increasing value of $\epsilon$, it is possible to observe the transition from weakly to strongly non-Markovian region ($\Phi_{t_2, t_1}$ breaking also P in addition of CP).

\begin{figure}[tb]
\begin{center}
\includegraphics[width=0.38\textwidth]{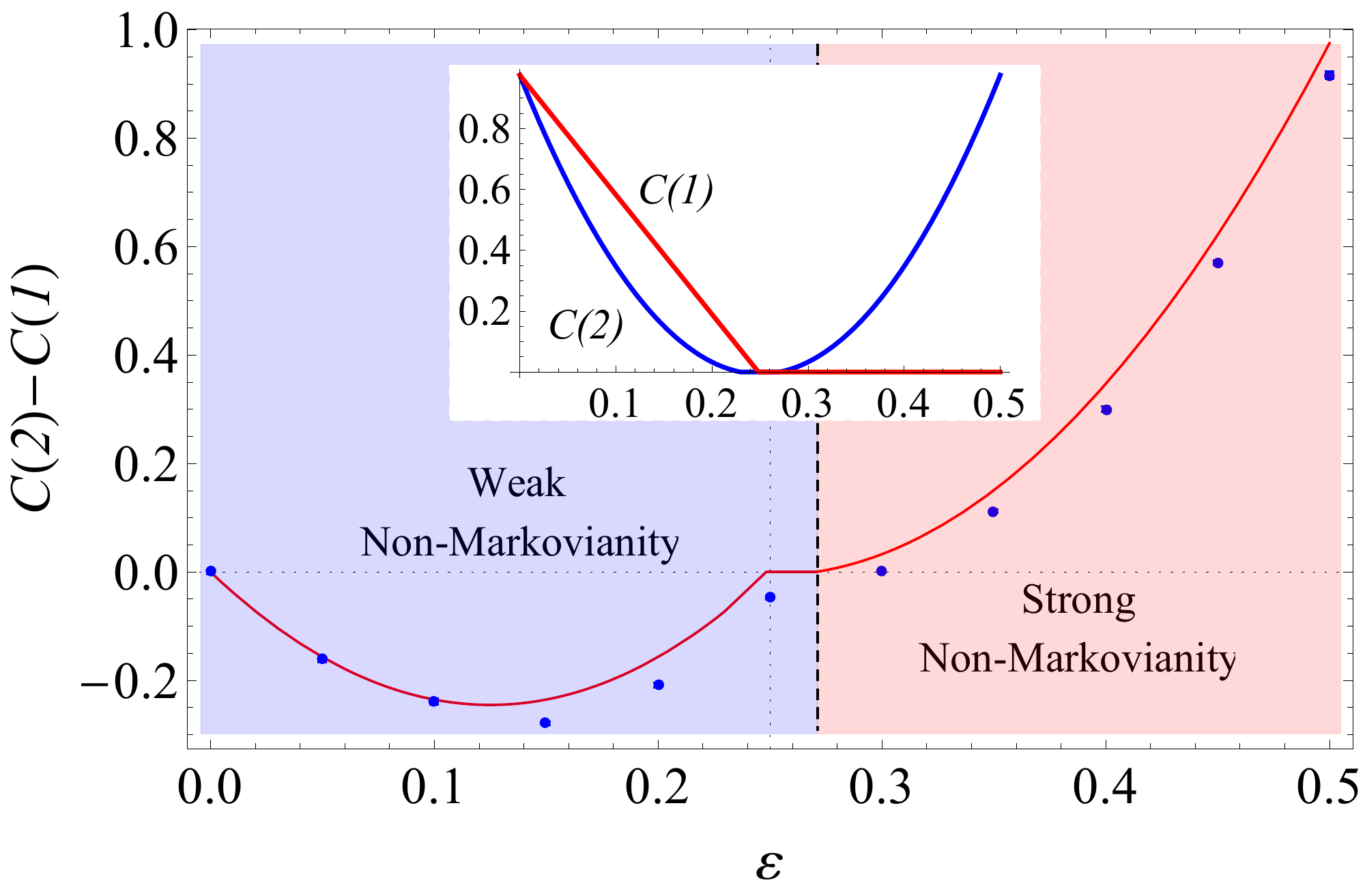}
\caption{\label{fig:2}
Transition from weak to strong non-Markovian regime with experimental control parameter $\epsilon$.
The entanglement $C(2)$ after the second collision is larger than entanglement $C(1)$ after the first collision when the dynamics belongs to strongly non-Markovian regime. The inset shows both the values of $C(1)$ and $C(2)$ as a function of the control parameter $\epsilon$.
Adapted from~\cite{Ber2015}.
}
\end{center}
\end{figure}

The experiment uses four liquid crystal (LC) cells to implement the collisions allowing also to tune the control parameter
  $\epsilon$~\cite{Ber2015}. 
 In particular, controlling the amount of time and timing sequence of applying the voltage to LC cells, allow the simulation of the  two collision dynamical map of Eq.~(\ref{eq:coll}) in different regimes. It is possible to show mathematically that the transition from weak to strong non-Markovian regime for single photon and for the considered dynamical map also coincides with the appearance of non-monotonic behaviour of polarization entanglement when ancillary photon is used. Here, the photon pair is prepared in fully entangled polarization state, and one of the photons goes through the collision sequence. Figure \ref{fig:2} shows the experimental result how much the entanglement (concurrence) changes in the second step of the protocol, i.e., $C(2)-C(1)$ where $C(2)$ [$C(1)$] is the amount of entanglement after step 2 (1). For small value of  $\epsilon$, the entanglement is reduced in the second collision  within the weakly non-Markovian regime. In the strongly non-Markovian regime, the entanglement revives in the second collision.  
In general, this is a proof-of-principle experiment allowing to detect different strengths of non-Markovianity with a conceptually simple and highly controllable set-up. It is worth mentioning that there also exists an experimental study on the spectral properties of dynamical maps and their relation to
non-Markovian dynamics~\cite{syu}.

\subsection{Non-Markovianity in NV-center systems with ambient environment}
NV-centers  in diamonds have become popular physical systems, e.g., for quantum information purposes~\cite{MWD} and recently they have also been used to study non-Markovian open system dynamics~\cite{haa}. For NV-system, the decoherence sources and corresponding interactions are inherently part of the total system in contrast to photonic systems where one can control, in laboratory setting, the polarization-frequency interaction at will.

In a NV-system, the carbon lattice in diamond has a point defect consisting of nitrogen atom with adjacent vacancy. 
This has electronic spin, which can be used to construct a qubit. This, in turn interacts with nitrogen nuclear spin, $^{13}C$ spins, and possible other sources of decoherence which form the combined environment for the electronic spin qubit. 
Within a given parameter regime and approximations, the system-environment interaction Hamiltonian $H_{SE}$, between the electronic spin and its environment, can be written as
$H_{SE} = S_z A_\parallel I_z + H_R$.
Here, $S_z$  is the electron and $I_Z$ nitrogen nuclear spin operator, $A_\parallel$ is coupling strength, and $H_R$ describes the interactions with the rest of the environment including $^{13}C$ atoms.  For the electron spin, this Hamiltonian induces  dephasing dynamics whose features can be controlled by tailoring the initial state of the nuclear spin. 

In preparation of the environmental nitrogen nuclear spin, it is first polarized and then rotated with a radio frequency pulse by an angle $\phi$.
In other words, this allows to prepare different initial superposition states for the nuclear spin, which then controls the dephasing of the electron spin, and the corresponding dynamics of its magnitude of coherences. In the experiment and with the used initial state, the latter is directly related to the magnitude of the Bloch vector of the open system electronic spin. This, in turn, is directly proportional to the amount of fluorescence light by the electron spin which is experimentally measured in the read-out phase of a Ramsey-type set-up~\cite{haa}.

\begin{figure}[tb]
\begin{center}
\includegraphics[width=0.44\textwidth]{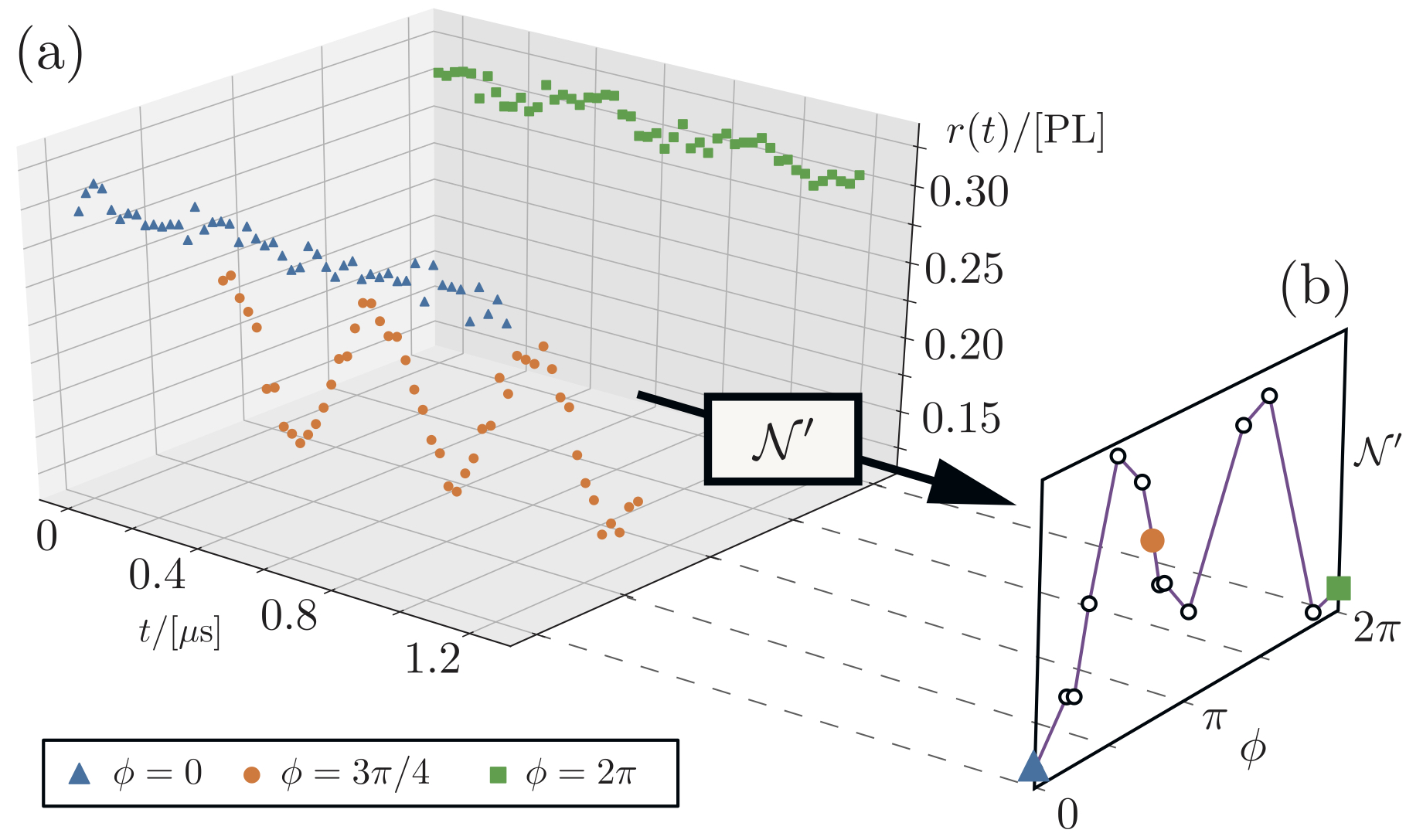}
\caption{\label{fig:3}
(a) The dynamics of magnitude of Bloch vector $r(t)$ for three values of experimental control parameter $\phi$.
 (b) Values of non-Markovianity $\mathcal{N'}$, for different values of parameter $\phi$, including the results of the three example cases presented in (a). Adapted from Fig.~S5 of~\cite{haa}.
}
\end{center}
\end{figure}

Figure~\ref{fig:3} (a) shows the dynamics of the magnitude of the Bloch vector for three different values of experimental control parameter 
$\phi$. In general, the amplitude of the oscillations is controlled by the initial state of the environmental nitrogen nuclear spin and the damping of oscillations, combined with reduction of Bloch vector magnitude, is provided by the rest of environment.
The results demonstrate  that for certain parameter values, the magnitude of the Bloch vector  behaves in non-monotonic manner indicating non-Markovian behaviour. This is seen in clear way in Fig.~\ref{fig:3} (b), which reports the values of non-Markovianity -- based on the trace distance measure with suitable modification for the current experimental purpose --  for several values of $\phi$.
The experiment contains a number of subtle points related, e.g., to uncontrollability of large part of the environmental state and accurate preparation 
of the nitrogen nuclear spin. Therefore, the analysis of full experimental data contains a sophisticated Bayesian model which allows to predict results also for unperformed measurements. This is another important and novel feature for open system studies,  for more details, see Fig.~3 of Ref.~\cite{haa}.
Note that NV centers have been used also in two other recent studies to study non-Markovian effects and for the control of entanglement dynamics~\cite{shp,fwa}.

\section{Applications of non-Markovian dynamics to quantum information protocols}

Apart from fundamental interest, it is also important to look for ways how non-Markovian memory-effects can be applied and exploited, e.g., in quantum information tasks. In this section we describe experimental examples for communication and computation purposes.

\subsection{Superdense coding}

In superdense coding Alice can send, in the optimal case, two bits of classical information by sending one qubit to Bob. By initially sharing an entangled Bell-state, Alice can navigate between the four Bell-states by applying local unitary - one of the four Pauli operations including identity - to her qubit. Thereby Alice is making a choice between four different options corresponding to two bits of classical information. By sending her qubit to Bob, who makes the Bell-measurement on the qubits, he reveals which one of the four choices Alice implemented. However, when the initial entanglement is decresed, e.g., because of noise and decoherence, the efficiency of the protocol is reduced. Applying non-Markovian noise and nonlocal memory effects allows to circumvent this problem~\cite{sdc,NLNM}.

Consider implemetation of the protocol with photons and suppose that before Alice receives her qubit, there is dephasing noise on her side
 due to polarization-frequency interaction according to Eq.~(\ref{eq:Hp})  producing  mixed bipartite polarization state.  Alice implements one of the four  local unitaries and sends her qubit to Bob. At this point, the bipartite polarization qubit state is mixed whilst the total polarization-frequency state is in a pure state. Before making the Bell measurement, Bob applies local dephasing noise, according to local polarization-frequency Hamiltonian Eq.~(\ref{eq:Hp}), to his original photon. If the initial two-photon frequency  probability distribution for photon 1 and 2, $|g(\omega_1, \omega_2)|^2$, does not contain correlations, i.e., $|g(\omega_1, \omega_2)|^2= |g(\omega_1)|^2  |g(\omega_2)|^2$, then Bob's noise reduces the efficiency of the protocol even further.
However, if the frequencies of the two photons are initially correlated  $|g(\omega_1, \omega_2)|^2\neq |g(\omega_1)|^2  |g(\omega_2)|^2$, and Bob applies the noise same amount of time as Alice did, then his noise allows to improve the efficiency of the transfer of information between Alice and Bob. 

\begin{figure}[tb]
\begin{center}
\includegraphics[width=0.35\textwidth]{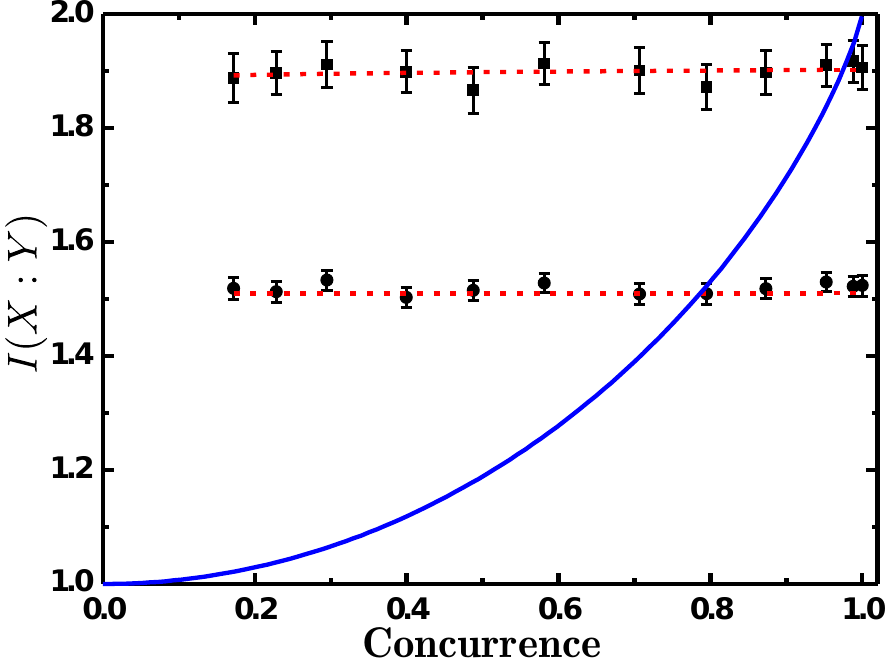}
\caption{\label{fig:4}
The efficiency of the superdense coding with nonlocal memory effects.
The mutual information as  a function of concurrence at the time when Alice makes the encoding.
The experimental results are displayed for 4-state (upper dots and line) and 3-state encoding (lower dots and line).
The solid curve describes the case when there is dephasing noise only on the side of Alice and no noise on Bob's side.
The efficiency of the protocol remains almost constant even though reducing the entanglement before Alice's encoding.
The figure is adapted from~\cite{sdc}.
}
\end{center}
\end{figure}

Considering bivariate Gaussian initial frequency distribution,
it is possible to show that the dense coding capacity takes the form  $C = 2-H\left(\frac{1+c_A^{2(1+K)}}{2}\right)\;.$ 
 Here, $c_A$ is the amount of shared qubit-qubit entanglement (concurrence) at the time when Alice makes the encoding, $K$ the correlation  coefficient between the frequencies of the two photons, and H is the binary entropy function. In particular having fully anti-correlated frequencies of the photons, $K=-1$, allows very efficient dense coding protocol even though the bipartite polarization state is practically fully mixed at the time when Alice makes her encoding. Experimental results displayed in Fig.~\ref{fig:4} shows that this is indeed the case. The mutual information, which gives the lower limit for channel capacity in this case, remains at high-value and almost constant when reducing the entanglement before Alice makes her encoding. 
This gives a proof-of-principle experimental demonstration on the usability of memory effects in a communication protocol.

 \subsection{Deutsch-Jozsa algorithm}

One of the first quantum algorithms, developed by Deutsch and Jozsa, is able to detect whether a given function is balanced or constant by single run of the quantum algorithm~\cite{dej}. In the refined Deutsch-Jozsa algorithm (RDJA) for single qubit~\cite{DJ,dco}, one applies a quantum black box transformation on the qubit corresponding to two possible unitaries -- and the question is whether measuring the qubit allows to detect which unitary was applied. 
This can be also interpreted as a  detection whether a coin is fake or fair by running the algorithm once.

Recently the single qubit refined  Deutsch-Jozsa algorithm was implemented with NV-centers in a diamond exploiting non-Markovian memory effects to 
enhance the success probability of the algorithm~\cite{DJ}. 
 Here, the nitrogen electron spin acts as a qubit used for implementing the protocol and nuclear spin represents the environment, in additon to $^{13}C$ spins causing decoherence.
The experiment is based on the following steps. After initializing the qubit to state $|0\rangle$, it is rotated with 
$(\pi/2)_x$ -- in general the notation $(\phi)_{x(y)}$ indicates rotation by angle $\phi$ around x(y) axis.
After this, phase gates are implemented. For the constant operations, either $U_1=(-\pi/2)_x (0)_y (-\pi/2)_x$ or $U_2=(-\pi/2)_x (2\pi)_y (-\pi/2)_x$
was used and for balanced operations  either $U_3=(-\pi/2)_x (3\pi)_y (-\pi/2)_x $ or $U_4=(-\pi/2)_x (\pi)_y (-\pi/2)_x$.
For read-out, another $(\pi/2)_x$ pulse is applied followed by population measurement in $\{|0\rangle, |1\rangle \}$ basis.
If the qubit is in the state  $\rho$, then the probability for $|0\rangle$ is $P_0=\tr (\rho |0\rangle\langle 0 |)$ and
for state $|1\rangle$ this is $P_1=1-P_0$. The probability of success of the protocol itself is  now the contrast between measured probabilities, e.g.,~$P_{0(U_3)}- P_{0(U_1)}$.

\begin{figure}[tb]
\begin{center}
\includegraphics[width=0.4\textwidth]{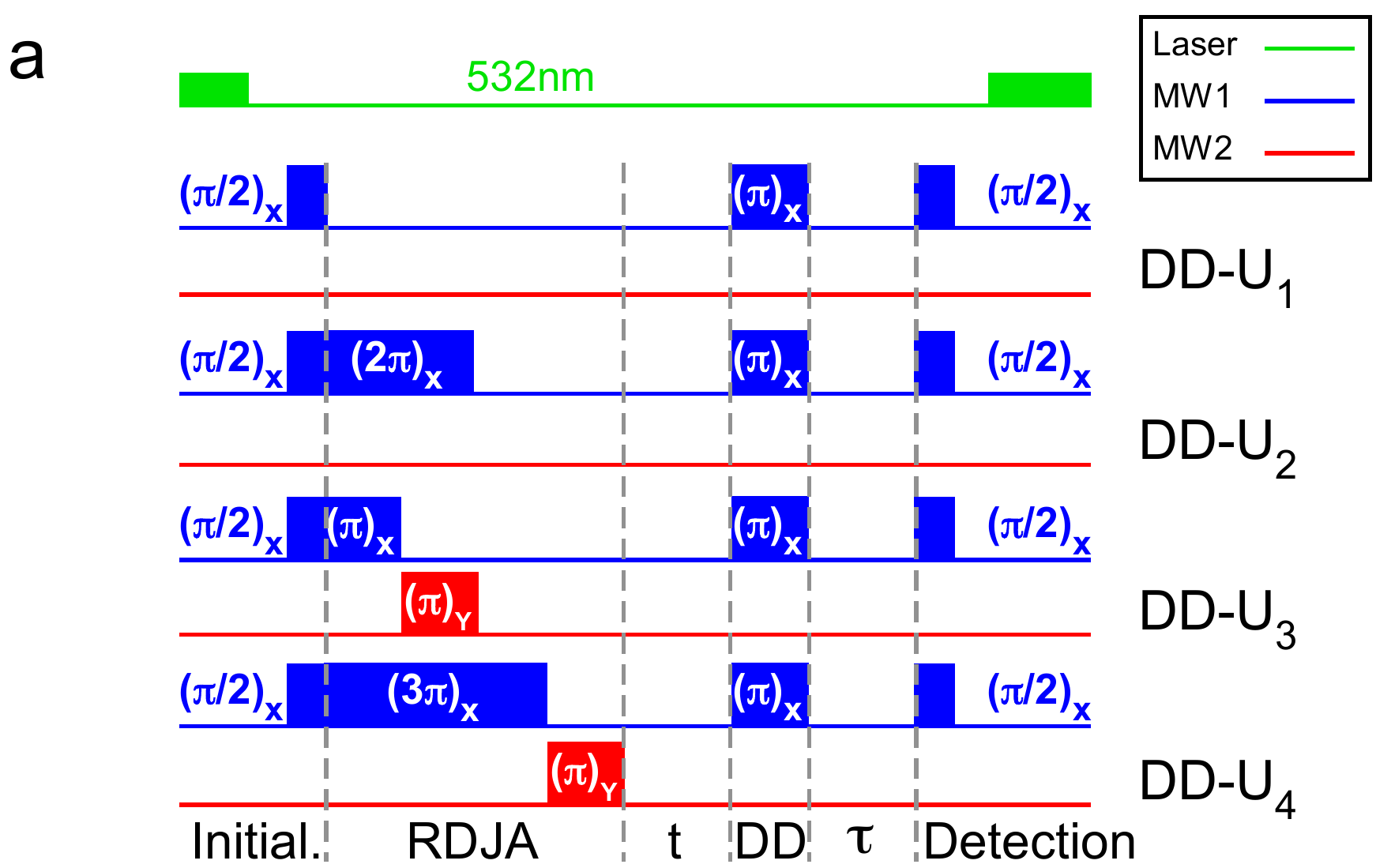}
\includegraphics[width=0.4\textwidth]{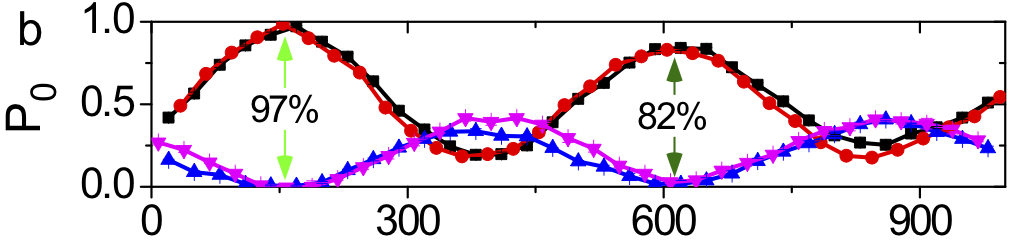}
\caption{\label{fig:5}
(a) Schematics of the initialization, RDJA protocol, and DD pulse.
(b) Probability $P_0$ as a function of the delay time $\tau$ (in units of ns) for
$U_3$ (blue upward triangles) and $U_4$ (cyan downward triangles) corresponding to balanced operations, and
$U_1$ (black squares) and $U_2$ (red dots) corresponding to constant operations.
Here $t=170$ and with a choice of $\tau \simeq 150$ns, the success probability of the RDJA algorithm reaches the value $0.97\%$.
The figure is adapted from~\cite{DJ}.
}
\end{center}
\end{figure}

 It takes finite amount of time to implement the algorithm and the presence of the intrinsic environment and subsequent decoherence reduces the probability of success of the algorithm.  
Now an interesting question is whether one should implement the read-out immediately after running the algorithm (phase gates), or delay it for certain time interval. The experimental results show that with non-Markovian environment one should indeed delay the read-out, when the algorithm has been already run, to improve the efficiency of the algorithm~\cite{DJ}. As a matter of fact, in the current case the most efficient implementation is obtained when non-Markovian memory effects are combined with dynamical decoupling  (DD) pulse.
Consider now the following scheme: i)  after running the algorithm, wait time $t$ before applying 
the dynamical decoupling pulse ii) after this, wait time $\tau$ and then implement the read-out of the qubit state.
For presentation of the protocol and all the pulse sequences, see Fig.~\ref{fig:5}(a).
The experimental results for the probabilities $P_0$ as a function of delay time $\tau$, for the choice $t=170$ns, are shown in Fig.~~\ref{fig:5} (b).  Probabilities oscillate -- and so does their contrast which gives the success probability of the protocol.
This reaches a very high value $0.97\%$ when $\tau  \simeq 150$ns and having   $t=170$ns. 
 Despite of the conceptual simplicity of the experiment, the results give a proof-of-principle demonstration that non-Markovian memory effects can be exploited when implementing quantum algorithms.

\section{Non-Markovian simulators and beyond Markovian -- non-Markovian discussion}
We are also interested in asking what type of non-Markovian noise processes in general - classical or quantum - can be implemented and simulated with open quantum systems. For the former, there was a recent experiment demonstrating how dephasing caused by randomly fluctuating external fields can be simulated with a photon~\cite{scia1}. Here, the implementation was done for random telegraph noise and Ornstein-Uhlenbeck stochastic process. The results demonstrated, e.g., that photons can be used to average over a large number of realizations of the stochastic process -- at once --  with the tomography of the polarization qubit density matrix. As interesting the results were, they did not -- however -- contain genuine quantum dynamics. 

A simulator for generic qubit dephasing with genuine time-evolution, in turn, was realized with photons  in Ref.~\cite{zdliu}.
The aim is to have full control of the dephasing dynamics, i.e.,~implementing arbitrary forms of the decoherence function.
Consider now the following initial polarization-frequency state for a photon.
$\ket{\Psi} = C_V\ket{V}\int  g(\omega)\ket{\omega}d\omega + C_H\ket{H}\int  e^{i\theta(\omega)} g(\omega)\ket{\omega}d\omega$.
 Note the frequency dependent phase factor $\theta(\omega)$ in the latter term.
Now the decoherence function of the dephasing dynamics becomes
\begin{equation}
 \kappa(t)=\int \vert g(\omega)\vert^2e^{i\theta(\omega)}e^{i 2 \pi \Delta n \omega t}  d\omega
 \end{equation}
 where the combination of engineering both the frequency probability disrtibution $ \vert g(\omega)\vert^2$ and initial phase factor $\theta (\omega)$ allow almost arbitrary freedom in implementing dephasing. Subsequently, photon's polarization state dynamics can be used to emulate the dephasing in large number of other physical systems. 

\begin{figure}[tb]
\begin{center}
\includegraphics[width=0.24\textwidth]{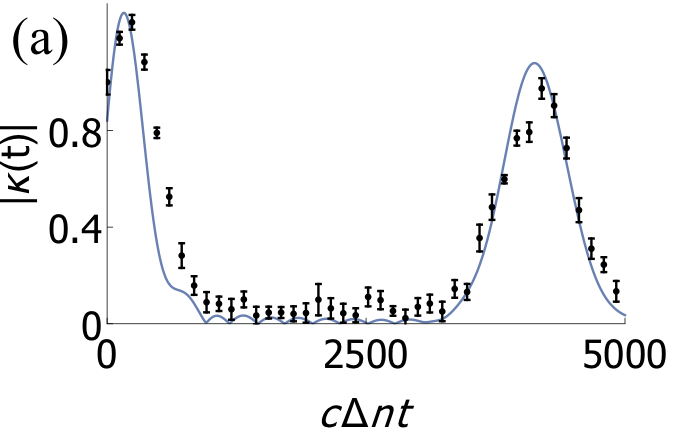}
\includegraphics[width=0.23\textwidth]{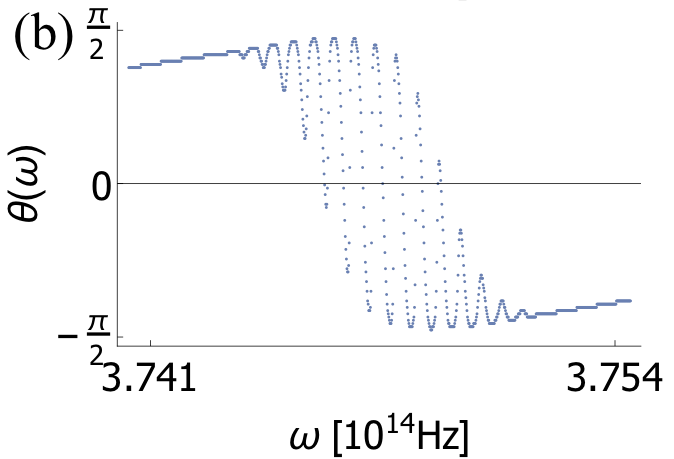}
\caption{\label{fig:6}
(a) Magnitude of decoherence function as a function of time when using a photon to simulate the influence of spin environment for a qubit. 
Time is measured as effective path difference. (b) The experimentally implemented corresponding frequency dependent phase distribution $\theta(\omega)$. For corresponding initial photon frequency probability distribution $ \vert g(\omega)\vert^2$, see Fig.~3 (a) of~\cite{zdliu}.
}
\end{center}
\end{figure}

In the experiment, gratings are used to convert frequency modes to spatial ones. Subsequently, spatial light modulator (SLM) is used to implement the required phase distribution 
$\theta(\omega)$ and the hologram of the SLM is used to modify the frequency probability distribution $ \vert g(\omega)\vert^2$.  Once the initial state is prepared, the dephasing time-evolution is implemented with birefringent quartz plates following the interaction Hamiltonian in Eq.~(\ref{eq:Hp}). 

To demonstrate the feasability of the set-up, the experimental results demonstrate, e.g.,  how to use photon to simulate the dephasing dynamics of a qubit coupled to Ising spin chain acting as environment. Figure \ref{fig:6} (a) shows the results for the dephasing dynamics of the qubit corresponding to spin environment being in paramagnetic phase. For the optical implementation, the corresponding frequency dependent initial phase distribution $\theta(\omega)$ is displayed in Fig.~\ref{fig:6} (b). For more details, see~\cite{zdliu}. Note that the set-up also allows an emulation of a non-positive dynamical map and also introduced the concept of synthetic spectral density, i.e., dephasing dynamics corresponding to spectral densities which would not otherwise appear in natural physical systems.   

\section{Conclusions and outlook}
The studies of open quantum systems are important both for fundamental reasons and for applications. 
Recent developments on controlling and engineering open systems, and their environments,  have allowed the 
experimental implementation and testing of several theoretical results of non-Markovian quantum dynamics. 
This include, e.g., controlling Markovian to non-Markovian transition~\cite{NMNP}, detecting various features of non-Markovian dynamics~\cite{Ber2015}, and mathematical properties 
of dynamical maps~\cite{syu}. For practical purposes and applications, the first proof-of-principle experiments demonstrate how non-Markovian memory effects 
can be exploited, e.g., to computation~\cite{DJ} and communication~\cite{sdc} purposes within the quantum domain.

Despite of their importance for proof-of-principle type of demonstrations, major part of recent experiments deal with conceptually rather simple open system dynamics. In many experimental considerations, one studies single qubits, and often their non-dissipative dynamics.
Therefore, there is a clear need for experimental studies on non-Markovian dynamics with complicated -- or even complex -- open systems by increasing the Hilbert space dimension and including also interaction within the subsystems. At the moment, not very much is known, e.g., how interactions among multi-partite open systems influence their non-Markovianity. Combining this with dissipative dynamics and genuine quantum baths in non-Markovian dynamics poses important challenges for future experimental work.There are also other areas, such as quantum metrology~\cite{awc} or probing of complex quantum systems~\cite{ph}, where interesting theoretical results exists though experimental implementations are still lacking to large extent. Note that the work on non-Markovian dynamics has opened also experimental possibilities for the local detection of quantum correlations~\cite{mge,jst}. 

While there has already been experimental demonstration of a quantum simulator for Markovian dynamics with trapped ions~\cite{jtbar,pschin},
multipurpose simulator for non-Markovian dynamics is still missing. Recent experimental results allow essentially arbitrary control of qubit dephasing dynamics~\cite{zdliu} but this has limitations, e.g., when going beyond dephasing and to dissipative systems. Perhaps, the use of IBM Q Experience open new avenues for this direction~\cite{gga}. Most of the experiments have been done for finite size discrete quantum systems. Also continuous variable (CV) open systems, such as quantum harmonic oscillators, are ubiquitous in physics. Though there exist early experimental results in this context for non-Markovian features of the open system dynamics~\cite{sgo}, quantum simulators for this purpose have not yet been experimentally realized. Here, there exists also an early theoretical proposal for simulating quantum Brownian motion with a single trapped ion~\cite{qss}.
Moreover, complex quantum networks in the context of optical multimode platform open also promising future directions for simulating CV open systems~\cite{paris}.
In general, we expect that both fundamental and applicative studies on non-Markovian quantum dynamics provide stimulating and interesting research problems for increasing number of experimental platforms in the future.

\acknowledgments
This work was supported by the National Key Research and Development Program of China (No. 2017YFA0304100), the National Natural Science Foundation of China (Nos. 11774335, 11821404), Key Research Program of Frontier Sciences, CAS (No. QYZDY-SSW-SLH003), the Fundamental Research Funds for the Central Universities (No. WK2470000026), and Anhui Initiative in Quantum Information Technologies (AHY020100).
We thank N.~K.~Bernardes, J.~F.~Haase, S.~Huelga and F.-W.~Sun for discussions.

\end{document}